\documentclass{article}

\usepackage[english]{babel}

\usepackage[letterpaper,top=2cm,bottom=2cm,left=3cm,right=3cm,marginparwidth=1.75cm]{geometry}

\usepackage{natbib}

\usepackage{junwei}
\usepackage{amsmath}
\usepackage{amsfonts}
\usepackage{amssymb}
\usepackage{graphicx}
\usepackage{csquotes}
\usepackage{bm}
\usepackage[colorlinks=true, allcolors=blue]{hyperref}
\usepackage{authblk}

\author[1]{Tianjun Ke$^*$}
\author[1]{Zhiyu Xu$^*$}

\affil[1]{Department of Statistics, Columbia University}

\title{Beyond Asymptotics: Practical Insights into Community Detection in Complex Networks}
\date{}

\begin{document}
\maketitle

\def\thefootnote{*}\footnotetext{Equal contribution.}

\vspace{-20pt}

\begin{abstract}
The stochastic block model (SBM) is a fundamental tool for community detection in networks, yet the finite-sample performance of inference methods remains underexplored. We evaluate key algorithms—spectral methods, variational inference, and Gibbs sampling—under varying conditions, including signal-to-noise ratios, heterogeneous community sizes, and multimodality. Our results highlight significant performance variations: spectral methods, especially SCORE, excel in computational efficiency and scalability, while Gibbs sampling dominates in small, well-separated networks. Variational Expectation-Maximization strikes a balance between accuracy and cost in larger networks but struggles with optimization in highly imbalanced settings. These findings underscore the practical trade-offs among methods and provide actionable guidance for algorithm selection in real-world applications. Our results also call for further theoretical investigation in SBMs with complex structures. The code can be found at \url{https://github.com/Toby-X/SBM_computation}.
\end{abstract}

\section{Introduction}
The idea of using networks to find latent community structures has been ubiquitous in biology \citep{chen2006detecting, marcotte1999detecting}, sociology \citep{fortunato2010community} and machine learning \citep{linden2003amazon}. 
Among various approaches, the stochastic block model (SBM, \citealp{holland1983stochastic}) stands out as the simplest generative model for identifying community structures. In undirected networks, symmetric adjacency matrices are used to represent connections, where the $(i,j)$-th entry is $1$ if and only if there is an edge between node $i$ and node $j$ and $0$ otherwise. For an undirected network with $N$ nodes and $K$ latent communities, SBM assumes that the expectation of the observed network has the following structure,
\begin{equation*}
    \label{eqn:def_sbm}
    \Ab^* := \mathbb{E} \mathbf{A} = \mathbf Z \mathbf B \mathbf Z^\top,
\end{equation*}
where $\Ab\in \{0,1\}^{N\times N}$ is the observed adjacency matrix, $\Zb \in \{0,1\}^{N\times K}$ is the community structure matrix, and $\Bb\in [0, 1]^{K \times K}$ is a symmetric matrix whose $(k,\ell)$-th entry represents the probability that there is an edge between community $k$ and community $\ell$. Each row of $\Zb$ contains exactly one entry equal to $1$, reflecting that each node belongs to precisely one community. 

The popularity of SBM lies in its sharp information-theoretic phase transitions for statistical inference, making it a useful tool for understanding the limits of information extraction in networks and evaluating algorithmic efficiency for community detection. Recent theoretical advancements in SBM have been extensively reviewed by \citet{abbe2018community}. Notably, \citet{lei2024computational} shows that under some specific asymptotic regime, the information-theoretic threshold for exact community structure recovery is the same for polynomial-time algorithms and the maximum likelihood estimator (MLE) for bipartite networks, with easy extensions to networks with $K$ latent communities. These results suggest that computationally efficient polynomial-time methods are sufficient for SBMs, obviating the need for more computationally intensive approaches. Despite these advances, several gaps remain. First, the theoretical results are confined to asymptotic settings, leaving the finite-sample performance of these algorithms unclear. Second, it remains uncertain which polynomial-time method—such as Approximate Message Passing \citep{feng2022unifying} or Spectral Clustering \citep{von2007tutorial}—is most effective in practice. Factors such as sample size, the number of latent communities, community size imbalance, and inter-community connectivity can significantly impact algorithmic performance, yet their impact remains underexplored.

To address this gap, we conduct a comprehensive investigation into the finite-sample performance of various statistical inference methods for community detection in SBMs across a wide range of challenging scenarios.  Specifically, we evaluate performance across extensive signal-to-noise ratios (SNRs), heterogeneous community sizes, and multimodal connectivity structures. Our findings provide actionable insights into the practical applicability of these algorithms while highlighting gaps in the theoretical understanding of finite-sample performance, particularly in the presence of noisy or unbalanced data.

\section{Methods}
In this section, we briefly discuss the methods considered in this paper. Spectral methods, being polynomial-time algorithms, are the most computationally efficient. Variational methods follow in terms of computational cost, while Gibbs sampling requires the most intensive computation.

\subsection{Gibbs Sampling}
Gibbs sampling uses the standard hierarchical structure to specify the stochastic block models \citep{nowicki2001estimation, golightly2005bayesian, mcdaid2013improved}:
\begin{align*}
    \mathbf Z_{i,:} &\sim \text{Mul}(\bpi),\\
    B_{ij} &\sim \text{Beta}(a,b), \\
    \bpi &\sim \text{Dir}(\alpha_1,\dots, \alpha_K),\\
    A_{ij} \given \Zb, \Bb &\sim \text{Ber}\left(\Zb_{i,:}^\top\mathbf B \Zb_{j,:}\right).
\end{align*}
The derivation of conditional distributions can be found in Appendix \ref{sec:gibbs}.

\subsection{Variational Bayes}
We consider a general variational Bayes method from \citet{zhang2020theoretical}, where they consider the optimization problem:
\begin{align} \label{eq:vb}
\max _{{\mathcal{Q}} \in \mathcal{S}_{\mathrm{MF}}} F(Q) := \max _{{\mathcal{Q}} \in \mathcal{S}_{\mathrm{MF}}} \int \log \phi_{\mathrm{Vec} (\Ab^*)}(\by) d Q(\bz, \Bb)-D\left(Q_{\bz} Q_{\Bb} Q_{\lambda} \| \Pi(\bz, \Bb)\right),
\end{align}
where $\mathcal{S}_{\mathrm{MF}}$ is the mean-field variational class and $\phi_{\btheta}(\by) := \frac{1}{(\sqrt{2\pi})^{n^2}} \exp \Big( -\frac{1}{2} \|\by - \btheta\|^2 \Big )$. We can obtain a variational algorithm for the stochastic block model with explicit update formulas thanks to the mean-field assumption. The complete description of the variational class and the algorithm can be found in Appendix \ref{sec:vb}.

\subsection{Variational-EM}
The variational Expectation-Maximization method (Variational-EM) \citep{leger2016blockmodels} considers 
\begin{gather*}
    \Zb  \sim \text{Mul}(\balpha), ~~~~
    \Ab_{ij} \given Z_{iq} Z_{j\ell} = 1 \sim \cF_{q\ell},
\end{gather*}
where $\balpha ^\top \bm{1} = 1$ and $\cF_{q\ell}$ is the model for community $q$ and $\ell$. A variational form of likelihood is considered as 
\begin{align*}
    J=\sum_{i, q} \tau_{i q} \log \left(\alpha_q\right)+\sum_{i, j ; i \neq j} \sum_{q, \ell} \tau_{i q} \tau_{j \ell} \log f_{q \ell}\left(A_{i j}\right),
\end{align*}
where $\btau_i$ is the variational parameters of the multinomial distribution which approximates $\Zb_{i,:} \given \Ab$. Details can be found in Appendix \ref{sec:vem}.

\subsection{Spectral Methods}
Spectral Clustering has been successfully applied to various fields for its simplicity and accuracy \citep{von2007tutorial}. In the context of the stochastic block model (SBM), the top $K$ eigenspace of the adjacency matrix exhibits a simplex structure, where each cluster corresponds to a vertex of the simplex \citep{chen2024spectral}. Building on this insight, spectral-based methods have been extended to degree-corrected stochastic block models \citealp{karrer2011stochastic}, leading to the development of approaches such as SCORE \citep{jin2015fast}, one-class SVM \citep{mao2018overlapping}, and regularized spectral clustering \citep{qin2013regularized}. Although these methods are designed to address broader scenarios than the standard SBM, it remains an open question whether they can effectively handle challenging settings such as community imbalance and sparsity. Given the diversity of spectral clustering techniques, we detail the specific methods considered in this paper and their terminology for clarity.

\paragraph{Spectral Clustering.} For vanilla spectral clustering, we first perform the top-$K$ eigenvalue decomposition of the adjacency matrix $\Ab$,
\begin{equation*}
    \Ab \approx \Ub\mathbf \Lambda\Ub^\top,
\end{equation*}
where $\Ub$ contains the top-$K$ eigenvectors, and $\mathbf \Lambda$ is the diagonal matrix with top-$K$ eigenvalues. Then, we apply K-means clustering to the rows of $\Ub$ to estimate $\Zb$.

\paragraph{SCORE.} SCORE normalizes the eigenvector matrix by dividing each column by the first column element-wise
\begin{equation*}
    \Ub_{\text{SCORE}} = \Ub/\Ub_{:,1}.
\end{equation*}
K-means clustering is then applied to $\Ub_{\text{SCORE}}$ to estimate $\Zb$.

\paragraph{$L_2$ Normalization.} In this method, each row of $\Ub$ is normalized by its $L_2$ norm:
\begin{equation*}
    \bu_{i,:}^{L_2} = \bu_{i,:}/\|\bu_{i,:}\|,\text{ i}\in [N],
\end{equation*}
where $\bu_{i,:}$ is the $i$-th row of $\Ub$ . The resulting normalized matrix, denoted as $\Ub_{L_2}$, is then clustered using K-means to estimate $\Zb$.

\paragraph{Regularized Spectral Clustering.}
This method constructs a regularized graph Laplacian:
\begin{equation*}
    \Lb_{\tau} = \Db_{\tau}^{-1/2}\Ab \Db_{\tau}^{-1/2},
\end{equation*}
where $\Db_{\tau}=\Db+\tau\Ib$, $\Db$ is a diagonal matrix with $\Db_{ii}=\sum_j\Ab_{i,j}$, and $\tau$ is set to be $\sum_{i=1}^N\Db_{ii}$ as suggested in \citet{qin2013regularized}.

\section{Numerical Experiments}

\subsection{Simulation Settings}

We consider a broad range of scenarios to evaluate the performance of our methods. For the number of nodes, we use $N = 250, 500, 1000, 2000$, representing moderate-sized graphs to larger ones. For the number of communities, we set $K = 5, 10, 20$, ranging from simpler structures to more complex configurations. Notably, even in a balanced dataset, the combination of $N = 250$ and $K = 20$ poses a significant challenge due to the limited information available for distinguishing a large number of diverse communities. To introduce heterogeneity in the cluster sizes, we adopt the approach outlined in \citep{zhang2022heteroskedastic}. Specifically, let $\balpha = (\alpha_1, \dots, \alpha_K)$, where 
$$
v_1, \dots, v_K \stackrel{\text{i.i.d.}}{\sim} \text{Uniform}(0,1), \quad \alpha_k = \frac{v_k^\beta}{\sum_{i=1}^K v_i^\beta}.
$$
Here, $\beta$ is a heterogeneity parameter controlling the imbalance in cluster sizes. A larger $\beta$ results in a greater imbalance in $\balpha$. We set $\beta = 0, 5, 10$ in our experiments to explore varying degrees of heterogeneity. For the kernel probability matrix, we follow a standard assumption in the literature \citep{lei2024computational}. The elements of the kernel matrix $\Bb$ are given by
$$
B_{ij} = 
\begin{cases} 
\frac{3}{2}\rho, & \text{if } i = j, \\ 
\frac{1}{2}\rho, & \text{otherwise},
\end{cases}
$$
where $\rho \in (0, 2/3)$. We note that $\rho = N^{-1}$ achieves exact recovery in asymptotic theory \citep{lei2024computational}. To evaluate performance under varying levels of information, we select $\rho = N^{-1}, N^{-0.5}, N^{-0.1}$. These values allow us to explore scenarios ranging from weak to strong signal levels, with $N^{-1}$ serving as the threshold for perfect recovery. However, perfect recovery may not always be attainable in finite samples, which is the focus of this study. Specific initialization can be found in Appendix \ref{sec:init}.

\subsection{Results}

\begin{figure}[th]
    \centering
    \includegraphics[width=1\linewidth]{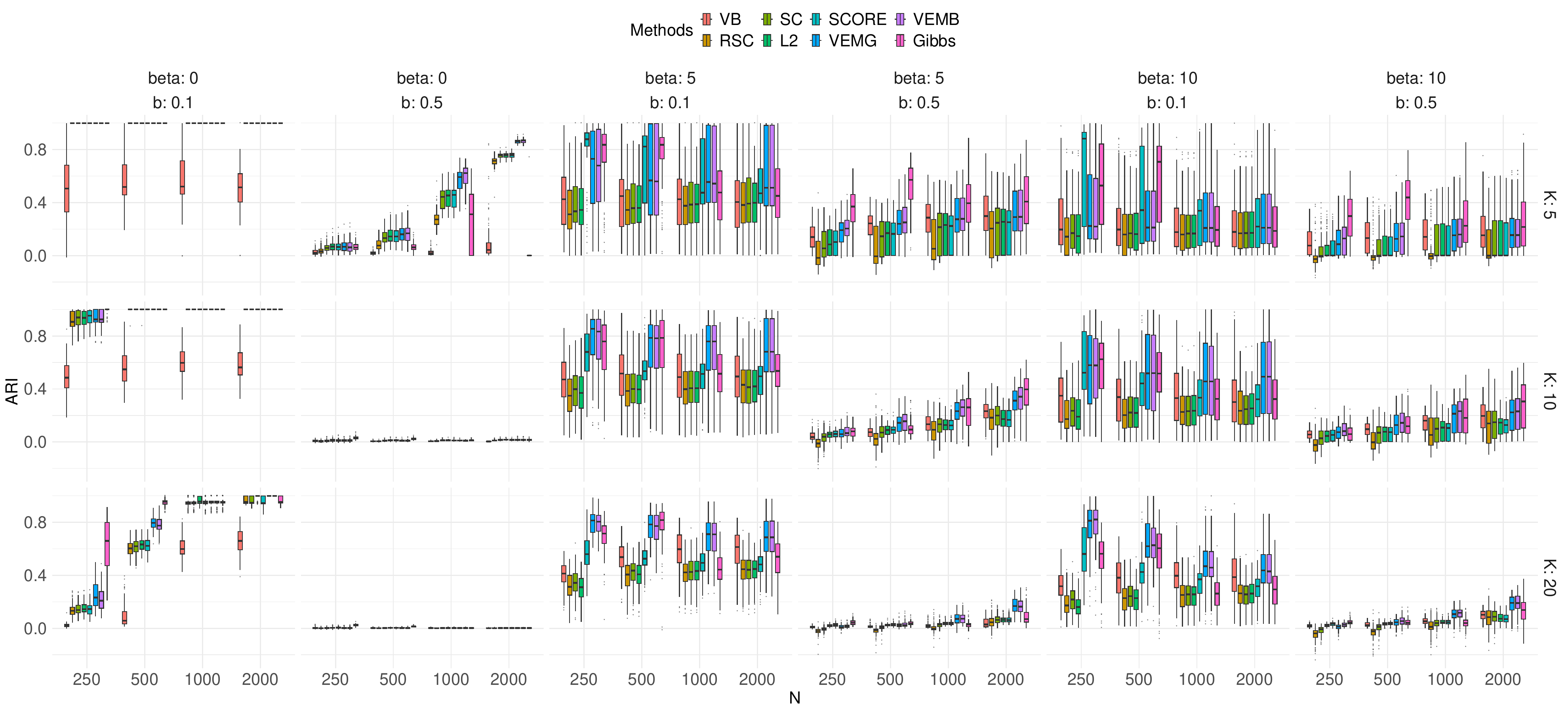}
    \caption{ARI comparisons of all methods with 100 different random seeds. L2 denotes the one-class SVM method. VEMB and VEMG denote variational-EM with the Bernoulli model and Gaussian model respectively. Other methods are represented by abbreviation. SCORE dominates other spectral methods with RSC performing the worst. Variational Bayes has inferior performance compared to variational-EM methods. Variational-EM deteriorates in performance as $N$ increases in challenging scenarios. Gibbs sampling only dominates in simple networks despite being an exact method.}
    \label{fig:ARI}
\end{figure}

In this project, we evaluate the performance of clustering algorithms using the adjusted Rand index (ARI; \citealp{hubert1985comparing}) and normalized mutual information (NMI; \citealp{strehl2002cluster}) as metrics. ARI measures how many samples are in the same cluster to see how different the clustering result is from the true value, with range $[-1, 1]$. Higher ARI values indicate more accurate clustering. As ARI and NMI yield similar results for our experiments, we focus our analysis of ARI, with results of NMI provided in Appendix \ref{sec:NMI}.

The results, shown in Figure \ref{fig:ARI}, reveal that the performance of the chosen inference methods varies across scenarios. Note that all methods fail in the sparsest case ($b=1$) so we exclude it in the figure. This highlights a dichotomy between finite-sample performance and the asymptotic guarantees discussed in \citet{lei2024computational}. We hypothesize that achieving the asymptotic information-theoretic limit requires significantly larger sample sizes.

One of the most peculiar phenomena occurs when $K=10$ or $20$ and $b=0.5$. In this case, the performances of all algorithms improve when $\beta$ increases to $5$ while all methods fail when $\beta=0$. On the one hand, since $\beta=0$ represents a balanced clustering size scenario, this failure likely arises because the sample sizes for individual clusters are too small to extract meaningful information. For instance, spectral clustering when $N=5000$ achieves ARI around $0.15$. This indicates increasing $\beta$ does not necessarily make the task harder, as the latent communities with a larger sample size can be learned more easily due to the imbalance. Hence, an improvement in ARI can be observed as we increase the imbalance of the communities.

Variational Bayes exhibits significantly worse performance under $\beta=0$ compared to other methods. In other scenarios, its performance typically falls between spectral methods and variational EM. We suspect that, under $\beta=0$, the evidence lower bound (ELBO) deviates more substantially from the posterior distribution, resulting in poor performance. Another explanation is that VB may have a slower convergence rate in the balanced scenario. Hence, we recommend using variational-EM instead of variational Bayes for similar computation complexity and better performance.

Among eigenvalue decomposition-based methods, regularized spectral clustering performs the worst, despite claims that it mitigates sparsity. The performances of vanilla spectral clustering and $L_2$ normalization are similar, while SCORE dominates. Although a vanilla SBM does not incorporate degree heterogeneity, using $L_2$ normalization and SCORE does not degrade performance. The superiority of SCORE may stem from its projection of $K$ dimensional eigenspace into a $K-1$ dimensional space, effectively improving the estimation in larger clusters by trading off smaller ones. This is especially evident in scenarios with $K=5$ and $b=0.1$ for $N=250$ or $500$, where SCORE consistently outperforms all other methods. In a nutshell, We recommend using $L_2$ normalization over vanilla spectral clustering for similar performance and its generalizability to degree-corrected stochastic block models. While SCORE dominates all spectral methods in our settings, it may run into stability issues in real-world datasets due to its unique normalization scheme, so further investigation is needed. Nonetheless, SCORE seems to be a plausible recommendation.

Variational EM (VEM) generally outperforms vanilla spectral clustering, despite its reliance on non-convex optimization. We suspect it results from the initialization of spectral clustering and further optimization is capable of finding an improvement of the local maximum near its starting point. Since spectral clustering gives robust results, VEM also gives robust results that are better than spectral clustering. Another interesting discovery of VEM is that in challenging scenarios ($\beta>0$ and $b>0.1$), VEM’s performance deteriorates with increasing $N$, both in terms of median accuracy and variation. We attribute this to two factors. First, the data generation mechanism we adopt in this project yields a drastically different set of data, especially when $N$ is large. As larger $N$ amplifies the variability in the data, we can expect an increase in the variation of the estimation accuracy. Furthermore, VEM may suffer from its optimization landscape when the data has a more complicated structure. The inability to find the global optimum of the nonconvex optimization problem may contribute to its performance deterioration.

We initially conjecture that Gibbs will dominate because variational methods use approximation that leads to error and spectral methods do not vitalize the likelihood information from the model. However, Gibbs outperforms only when both $K$ and $N$ are small. We propose two potential explanations for this phenomenon. First, when $K$ is large, Gibbs may struggle to sample equally from all modes in practice, leading to estimation bias. Second, when $N$ is large, we notice that the performance of Gibbs is only slightly worse than VEM. This may be attributed to the influence of an uninformative prior. As $K$ increases, the effective sample size for each parameter shrinks, giving the prior greater sway in the posterior and subsequently degrading Gibbs sampling performance compared to VEM.


\section{Conclusion}

In this work, we investigated the efficacy of various statistical methods for community detection in stochastic block models under noisy, heterogeneous, and multimodal conditions. Our findings highlight significant differences in performance across methods, emphasizing the importance of context when selecting an inference algorithm.

For small networks with simple structures, Gibbs sampling consistently outperformes other methods. For larger networks and communities, variational Expectation-Maximization achieves the best balance between computational cost and accuracy. In scenarios requiring scalability, spectral clustering emerged as a practical choice.

Future work could explore more effective initialization schemes for Gibbs and variational EM, especially using SCORE. Additionally, designing metrics or simulation settings that ensure proper discovery of all clusters may better evaluate these methods. Our work also calls for deeper theoretical analysis to discover all communities when imbalanced, where we conjecture there should be a lower bound for the sample size of the smallest community.

\bibliographystyle{ims}
\bibliography{main}

\begin{thebibliography}{22}
\expandafter\ifx\csname natexlab\endcsname\relax\def\natexlab#1{#1}\fi
\expandafter\ifx\csname url\endcsname\relax
  \def\url#1{\texttt{#1}}\fi
\expandafter\ifx\csname urlprefix\endcsname\relax\def\urlprefix{URL }\fi

\bibitem[{Abbe(2018)}]{abbe2018community}
\textsc{Abbe, E.} (2018).
\newblock Community detection and stochastic block models: recent developments.
\newblock \textit{Journal of Machine Learning Research} \textbf{18} 1--86.

\bibitem[{Chen and Yuan(2006)}]{chen2006detecting}
\textsc{Chen, J.} and \textsc{Yuan, B.} (2006).
\newblock Detecting functional modules in the yeast protein--protein interaction network.
\newblock \textit{Bioinformatics} \textbf{22} 2283--2290.

\bibitem[{Chen and Gu(2024)}]{chen2024spectral}
\textsc{Chen, L.} and \textsc{Gu, Y.} (2024).
\newblock A spectral method for identifiable grade of membership analysis with binary responses.
\newblock \textit{Psychometrika}  1--32.

\bibitem[{Feng et~al.(2022)Feng, Venkataramanan, Rush, Samworth et~al.}]{feng2022unifying}
\textsc{Feng, O.~Y.}, \textsc{Venkataramanan, R.}, \textsc{Rush, C.}, \textsc{Samworth, R.~J.} \textsc{et~al.} (2022).
\newblock A unifying tutorial on approximate message passing.
\newblock \textit{Foundations and Trends{\textregistered} in Machine Learning} \textbf{15} 335--536.

\bibitem[{Fortunato(2010)}]{fortunato2010community}
\textsc{Fortunato, S.} (2010).
\newblock Community detection in graphs.
\newblock \textit{Physics reports} \textbf{486} 75--174.

\bibitem[{Golightly and Wilkinson(2005)}]{golightly2005bayesian}
\textsc{Golightly, A.} and \textsc{Wilkinson, D.~J.} (2005).
\newblock Bayesian inference for stochastic kinetic models using a diffusion approximation.
\newblock \textit{Biometrics} \textbf{61} 781--788.

\bibitem[{Holland et~al.(1983)Holland, Laskey and Leinhardt}]{holland1983stochastic}
\textsc{Holland, P.~W.}, \textsc{Laskey, K.~B.} and \textsc{Leinhardt, S.} (1983).
\newblock Stochastic blockmodels: First steps.
\newblock \textit{Social networks} \textbf{5} 109--137.

\bibitem[{Hubert and Arabie(1985)}]{hubert1985comparing}
\textsc{Hubert, L.} and \textsc{Arabie, P.} (1985).
\newblock Comparing partitions.
\newblock \textit{Journal of classification} \textbf{2} 193--218.

\bibitem[{Jin(2015)}]{jin2015fast}
\textsc{Jin, J.} (2015).
\newblock {Fast community detection by SCORE}.
\newblock \textit{The Annals of Statistics} \textbf{43} 57 -- 89.

\bibitem[{Karrer and Newman(2011)}]{karrer2011stochastic}
\textsc{Karrer, B.} and \textsc{Newman, M.~E.} (2011).
\newblock Stochastic blockmodels and community structure in networks.
\newblock \textit{Physical Review E—Statistical, Nonlinear, and Soft Matter Physics} \textbf{83} 016107.

\bibitem[{Leger(2016)}]{leger2016blockmodels}
\textsc{Leger, J.-B.} (2016).
\newblock Blockmodels: A r-package for estimating in latent block model and stochastic block model, with various probability functions, with or without covariates.
\newblock \textit{arXiv preprint arXiv:1602.07587} .

\bibitem[{Lei et~al.(2024)Lei, Zhang and Zhu}]{lei2024computational}
\textsc{Lei, J.}, \textsc{Zhang, A.~R.} and \textsc{Zhu, Z.} (2024).
\newblock Computational and statistical thresholds in multi-layer stochastic block models.
\newblock \textit{The Annals of Statistics} \textbf{52} 2431--2455.

\bibitem[{Linden et~al.(2003)Linden, Smith and York}]{linden2003amazon}
\textsc{Linden, G.}, \textsc{Smith, B.} and \textsc{York, J.} (2003).
\newblock Amazon. com recommendations: Item-to-item collaborative filtering.
\newblock \textit{IEEE Internet computing} \textbf{7} 76--80.

\bibitem[{Mao et~al.(2018)Mao, Sarkar and Chakrabarti}]{mao2018overlapping}
\textsc{Mao, X.}, \textsc{Sarkar, P.} and \textsc{Chakrabarti, D.} (2018).
\newblock Overlapping clustering models, and one (class) svm to bind them all.
\newblock \textit{Advances in Neural Information Processing Systems} \textbf{31}.

\bibitem[{Marcotte et~al.(1999)Marcotte, Pellegrini, Ng, Rice, Yeates and Eisenberg}]{marcotte1999detecting}
\textsc{Marcotte, E.~M.}, \textsc{Pellegrini, M.}, \textsc{Ng, H.-L.}, \textsc{Rice, D.~W.}, \textsc{Yeates, T.~O.} and \textsc{Eisenberg, D.} (1999).
\newblock Detecting protein function and protein-protein interactions from genome sequences.
\newblock \textit{Science} \textbf{285} 751--753.

\bibitem[{McDaid et~al.(2013)McDaid, Murphy, Friel and Hurley}]{mcdaid2013improved}
\textsc{McDaid, A.~F.}, \textsc{Murphy, T.~B.}, \textsc{Friel, N.} and \textsc{Hurley, N.~J.} (2013).
\newblock Improved bayesian inference for the stochastic block model with application to large networks.
\newblock \textit{Computational Statistics \& Data Analysis} \textbf{60} 12--31.

\bibitem[{Nowicki and Snijders(2001)}]{nowicki2001estimation}
\textsc{Nowicki, K.} and \textsc{Snijders, T. A.~B.} (2001).
\newblock Estimation and prediction for stochastic blockstructures.
\newblock \textit{Journal of the American statistical association} \textbf{96} 1077--1087.

\bibitem[{Qin and Rohe(2013)}]{qin2013regularized}
\textsc{Qin, T.} and \textsc{Rohe, K.} (2013).
\newblock Regularized spectral clustering under the degree-corrected stochastic blockmodel.
\newblock \textit{Advances in neural information processing systems} \textbf{26}.

\bibitem[{Strehl and Ghosh(2002)}]{strehl2002cluster}
\textsc{Strehl, A.} and \textsc{Ghosh, J.} (2002).
\newblock Cluster ensembles---a knowledge reuse framework for combining multiple partitions.
\newblock \textit{Journal of machine learning research} \textbf{3} 583--617.

\bibitem[{Von~Luxburg(2007)}]{von2007tutorial}
\textsc{Von~Luxburg, U.} (2007).
\newblock A tutorial on spectral clustering.
\newblock \textit{Statistics and computing} \textbf{17} 395--416.

\bibitem[{Zhang et~al.(2022)Zhang, Cai and Wu}]{zhang2022heteroskedastic}
\textsc{Zhang, A.~R.}, \textsc{Cai, T.~T.} and \textsc{Wu, Y.} (2022).
\newblock Heteroskedastic pca: Algorithm, optimality, and applications.
\newblock \textit{The Annals of Statistics} \textbf{50} 53--80.

\bibitem[{Zhang(2020)}]{zhang2020theoretical}
\textsc{Zhang, F.} (2020).
\newblock \textit{Theoretical Guarantees of Variational Inference and Its Applications}.
\newblock Ph.D. thesis, The University of Chicago.

\end{thebibliography}

\newpage
\appendix
\section{Appendix}

\subsection{Initialization}
\label{sec:init}
For Gibbss sampling, we use $a=b=2, \ \alpha_i = 1$ for all $i = 1, \ldots, K$. Variational Bayes uses a uniform initialization of the memberships and a standard initialization for the variational parameters as specified in Algorithm \ref{algo:vb}. Variational-EM is initialized with spectral clustering for both models. We run each scenario with 100 different random seeds and report the results in the below section.

\subsection{NMI}
\label{sec:NMI}

\begin{figure}[th]
    \centering
    \includegraphics[width=1\linewidth]{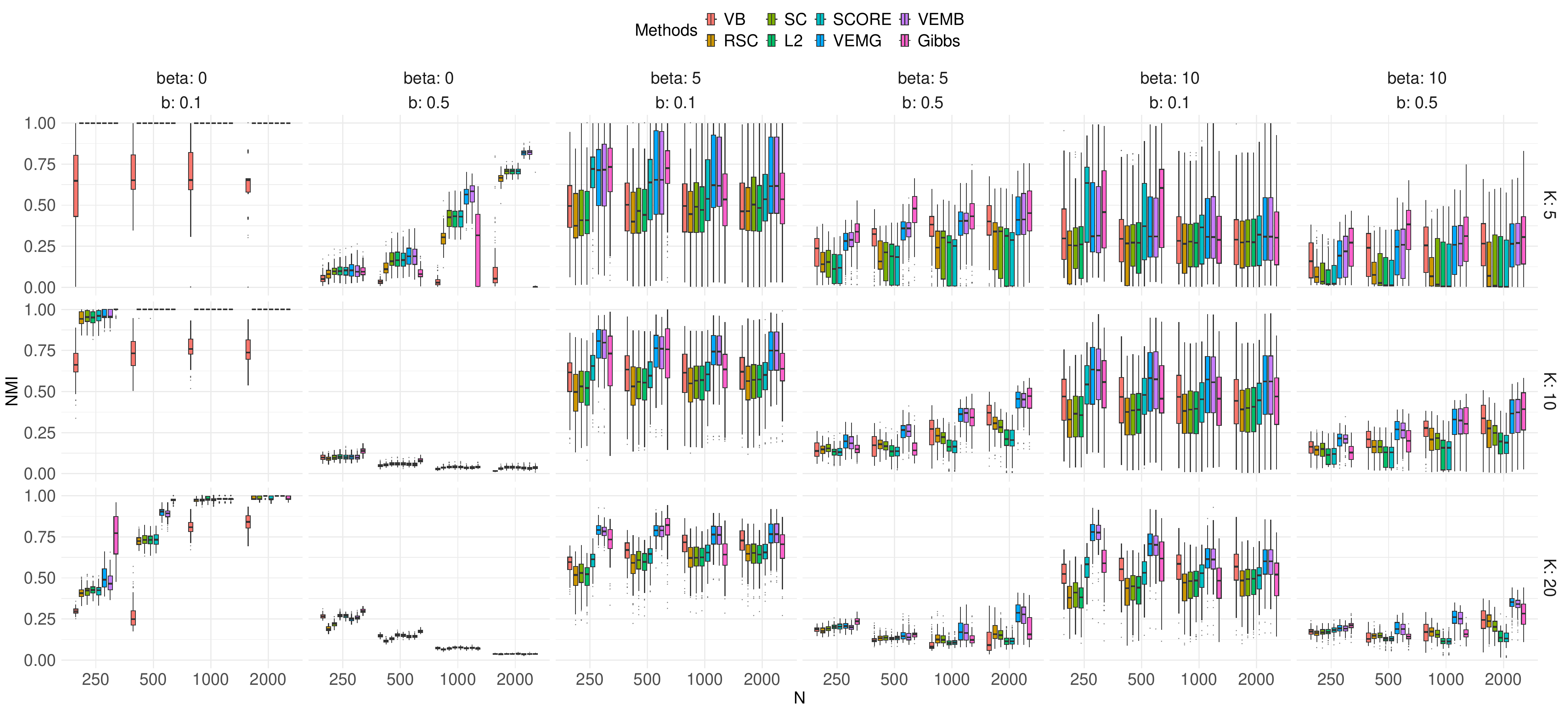}
    \caption{NMI comparisons of all methods with 100 different random seeds. L2 denotes the one-class SVM method. VEMB and VEMG denote variational-EM with the Bernoulli model and Gaussian model respectively. Other methods are represented by abbreviation. Results are similar to ARI.}
    \label{fig:NMI}
\end{figure}

\subsection{Gibbs Sampling}
\label{sec:gibbs}
Let
\begin{align*}
    n_k &= \sum_{i=1}^N\II(\Zb_{i,:}=\be_k),\quad k=1,2,\dots,K,\\
    n_{k\ell} &= \sum_{i\neq j}\II(\Zb_{i,:}=\be_k,\Zb_{j,:}=\be_\ell)=n_kn_\ell-n_k\II(k=\ell),\\
    A[k\ell] &= \sum_{(i,j):\Zb_{i,:}=\be_k,\Zb_{j,:}=\be_\ell}A_{ij},\quad k,s = 1,2,\dots,K.
\end{align*}
The full conditional distribution of each parameter can be derived as 
\begin{align*}
    \pi \given \cdot &\sim \text{Dir}(\alpha_1+n_1,\dots,\alpha_K+n_K),\\
    B_{k\ell} \given \cdot &\sim \text{Beta}(a+A[k\ell],1+n_{k\ell}-A[k\ell]),\\
    P(\Zb_{i, :}=\be_k \given \cdot) &\propto \pi_k\cdot \left(\prod_{j\neq i}B_{kz_j}^{A_{ij}}(1-B_{kz_j})^{1-A_{ij}} \right)\cdot \left(\prod_{j\neq i}B_{z_jk}^{A_{ji}}(1-B_{z_jk})^{1-A_{ji}} \right).
\end{align*}

\subsection{Variational Bayes} \label{sec:vb}

The mean-field variational class is specified as
\begin{align*}
\mathcal{S}_{\mathrm{MF}}= & \{Q: Q(\Zb, \Bb, \lambda)=Q_{\bz}(\bz) Q_{\Bb}(\Bb) Q_{\lambda}(\lambda),   \\
&~ \bz \in \overline{\mathcal{Z}}, \Bb \in \mathbb{R}^{k \times k}, \lambda \in \mathbb{R}, Q_{\Zb} \in \mathcal{S}_{\Zb}, Q_{\Bb} \in \mathcal{S}_{\Bb}, Q_{\lambda} \in \mathcal{S}_{\lambda}\},
\end{align*}
where $\overline{\mathcal{Z}} := \{\bz \in [k]^n: \sum_{i=1}^n \II \{z_i = t\} > 0, \forall t \in [k] \}$ is the support of $\bz$ and $\mathcal{S}_{\bz}, \mathcal{S}_{\Bb}, \mathcal{S}_{\lambda}$ are the variational families. In Equation (\ref{eq:vb}), $\mathcal{S}_{\bz}, \mathcal{S}_{\Bb}$ and $\mathcal{S}_{\lambda}$ respectively denote some distribution families for $\bz \in \overline{\mathcal{Z}}, \Bb \in \mathbb{R}^{k \times k}$ and $\lambda \in \mathbb{R}$. \citet{zhang2020theoretical} consider the following mass point distribution family for $\mathcal{S}_{Z}$:
$$
\mathcal{S}_{\zb}=\left\{Q_{\zb}: Q_{\zb}(\zb=\widetilde{\zb})=1, \text { for some } \widetilde{\zb} \in \overline{\mathcal{Z}}\right\}.
$$
In addition, $\mathcal{S}_{\Bb}$ is selected as the normal distribution family and $\mathcal{S}$ is selected as a parametric distribution family with parameter $a>0$. Here we abuse the notation a little bit and assume a vectorized $\mathrm{Vec}(\Bb)$ while we still use $\Bb$ to denote the vectorized matrix. The distribution families are defined as follows:
$$
\begin{gathered}
\mathcal{S}_{\Bb}=\left\{Q_{\Bb}: Q_{\Bb}=N(\bmu, \bSigma), \text { for } \bmu \in \mathbb{R}^{k^2}, \bSigma \in \mathbb{R}^{k^2 \times k^2}, \bSigma \succ 0\right\}, \\
\mathcal{S}_{\lambda}=\left\{Q_{\lambda}: \frac{d Q_{\lambda}}{d \lambda}=\sqrt{\frac{\beta}{\pi}} \lambda^{-3 / 2} \exp \left(\sqrt{2 a \beta}-\frac{\beta}{\lambda}-\frac{a \lambda}{2}\right), \text { for some } a>0\right\}.
\end{gathered}
$$
The complete algorithm can be obtained as in Algorithm \ref{algo:vb}.

\begin{algorithm}[th]
\caption{Variational Algorithm for Stochastic Block Model}
\begin{algorithmic}
\label{algo:vb}
\STATE \textbf{Input}: Number of clusters $K$, initial labels $\bz^{[0]} \in \overline{\mathcal{Z}}$, maximum iteration number $M$, tolerate level $\epsilon$.

\STATE \textbf{Initialize}: Iteration number $t=0$, objective function $L_{0}=\infty$, change of objective function $\nabla L_{0}=\infty, a^{[0]}=n^{2}$ and $\delta^{[0]}=1+\sqrt{\frac{2 \beta}{n^{2}}}$. The initial $\bmu$ and $\bSigma$ are computed by followings:
$$
\begin{gathered}
\mu_{cd}^{[0]}=\left(\delta^{[0]}\right)^{-1}\left(n_{c}^{[0]}\right)^{-1}\left(n_{d}^{[0]}\right)^{-1} \sum_{i=1}^{n} \sum_{j=1}^{n} A_{i j} \mathbb{I}\left\{z_{i}^{[0]}=c, z_{j}^{[0]}=d\right\}, \quad 1 \leq c, d \leq k ; \\
\Sigma_{c d}^{[0]}=\left(\delta^{[0]}\right)^{-1}\left(n_{c}^{[0]}\right)^{-1}\left(n_{d}^{[0]}\right)^{-1}, \quad 1 \leq c, d \leq k ;
\end{gathered}
$$
where $n_{c}^{[0]}=\sum_{i=1}^{n} \mathbb{I}\left\{z_{i}^{[0]}=c\right\}$.
\WHILE {$t<M$ and $\nabla L_{t}>\epsilon$} 
\STATE $t \leftarrow t+1$.
\STATE \textbf{Update} $Z$ :
\FOR {$i=1, \cdots, n$ in a randomized order}
\FOR {$c=1, \cdots, k$}
\STATE Set $z_{j}(i)^{[t]}$ to be the current label for $z_{j}$ and $n_{c}(i)=\sum_{j \neq i} \mathbb{I}\left\{z_{j}(i)^{[t]}=c\right\}$.
\STATE Compute
$$
\begin{aligned}
& v_{i c}=-k \log \left(1+n_{c}(i)^{-1}\right)-2 \sum_{j \neq i}^{n} A_{i j} \mu_{c z_{j}(i)^{[t]}}, \\
& +\delta^{[t-1]}\left[\sum_{j \neq i} \mu_{c z_{j}(i)^{[t]}}^{2}+\frac{1}{2} \mu_{c c}^{2}\right]+\frac{1}{2}\left(\sum_{r=1}^{k} \frac{n_{r}(i)}{n_{r}^{[t-1]}}+\frac{1}{n_{c}^{[t-1]}}\right)^{2}.
\end{aligned}
$$
\ENDFOR
\STATE Set $z_{i}^{[t]}=\operatorname{argmin}_{1 \leq c \leq k}\left\{v_{i c}\right\}$.
\ENDFOR
\STATE \textbf{Update} $a$ \textbf{and} $\delta$ : compute $n_{c}^{[t]}=\sum_{i=1}^{n} \mathbb{I}\left\{z_{i}^{[t]}=c\right\}$, then
$$
a^{[t]}=\sum_{i=1}^{n} \sum_{j=1}^{n}\left(\mu^{[t-1]}_{ z_{i}^{[t]} z_{j}^{[t]}}\right)^{2}+\left(\delta^{[t-1]}\right)^{-1}\left(\sum_{c=1}^{k} \frac{n_{c}^{[t]}}{n_{c}^{[t-1]}}\right)^{2}, \quad \delta^{[t]}=1+\sqrt{\frac{2 \beta}{a^{[t]}}}.
$$
\STATE \textbf{Update} $\mu$ \textbf{and} $\Sigma$:
$$
\begin{gathered}
\mu_{c d}^{[t]}=\left(\delta^{[t]}\right)^{-1}\left(n_{c}^{[t]}\right)^{-1}\left(n_{d}^{[t]}\right)^{-1} \sum_{i=1}^{n} \sum_{j=1}^{n} A_{i j} \mathbb{I}\left\{z_{i}^{[t]}=c, z_{j}^{[t]}=d\right\}, \quad 1 \leq c, d \leq k \\
\Sigma_{c d}^{[t]}=\left(\delta^{[t]}\right)^{-1}\left(n_{c}^{[t]}\right)^{-1}\left(n_{d}^{[t]}\right)^{-1}, \quad 1 \leq c, d \leq k
\end{gathered}
$$
\STATE \textbf{Update} $L_{t}$ \textbf{and} $\nabla L_{t}:$
$$
\begin{gathered}
L_{t}=\frac{k(k+1)}{4} \log \frac{\delta^{[t]}}{4 \beta e^{2}}+\sqrt{\frac{a^{[t]} \beta}{2}}-\frac{1}{2} \sum_{i=1}^{n} \sum_{j=1}^{n} A_{i j} \mu_{z_{i}^{[t]} z_{j}^{[t]}}^{[t]}+(D+1)\left(\frac{1}{2} k(k+1)+n \log k\right), \\
\nabla L_{t}=\left|L_{t-1}-L_{t}\right|.
\end{gathered}
$$
\ENDWHILE
\STATE \textbf{Output}: $\widehat{z}^{(k)}=z^{[t]}, \widehat{a}^{(k)}=a^{[t]}, \widehat{\mu}^{(k)}=\mu^{[t]}, \widehat{\Sigma}^{(k)}=\Sigma^{[t]}, \widehat{L}^{(k)}=L_{t}$.
\end{algorithmic}
\end{algorithm}

\subsection{Variational-EM} 

\label{sec:vem}

The EM procedure is described as follows:
\begin{enumerate}
    \item E step: Maximization with respect to variational parameters $\btau$.
    \item M step: Maximization with respect to original parameters $\balpha$ and function parameters in $\cF$.
\end{enumerate}
We consider Bernoulli and Gaussian as the function models, as following. For both models, we select the community with maximized value as the membership for each node.
\subsubsection{Bernoulli}
This is the common stochastic block model. The model is defined as follows:
\begin{equation*}
    \cF_{q\ell} = \text{Ber}(\pi_{q\ell}),
\end{equation*}
where $\pi_{q\ell} \in [0, 1]$ for $q, \ell = 1, \ldots, K$.

\subsubsection{Gaussian}
This model assumes a normal distribution of the community links, defined as:
\begin{equation*}
    \cF_{q\ell} = \cN(\mu_{q \ell}, \sigma^2),
\end{equation*}
where $\mu_{q \ell} \in \RR$ and $\sigma^2 \in \RR^+$ for $q, \ell = 1, \ldots, K$.

\end{document}